\documentclass[namedreferences]{solarphysics}
\usepackage{txfonts}
\usepackage[optionalrh]{spr-sola-addons} 
\usepackage{graphicx}        
\usepackage{color}           
\usepackage{url}             

\newcommand{\etal}{{\it et al.}}

\newcommand{\rmd}{ {\ \mathrm d} }

\newcommand{\apj}{    {\it Astrophys. J.}}

\newcommand{\caa}{   {\it Chin. Astron. Astrophys.}}

\newcommand{\pasp}{   {\it Pub. Astron. Soc. Pac.}}

\newcommand{\raa}{    {\it Res. Astron. Astrophys.}}
\newcommand{\solphys}{{\it Solar Phys.}}

\begin{document}

\begin{article}

\begin{opening}

\title{Using a New Sky Brightness Monitor to Observe the Annular Solar Eclipse on 15 January 2010}

\author{Y.~\surname{Liu}\sep
        Y.-D.~\surname{Shen}\sep
        X.-F.~\surname{Zhang}\sep
        N.-P.~\surname{Liu}
       }
\runningauthor{Liu et al.} \runningtitle{Using a New SBM to Observe an Annular Solar Eclipse}

   \institute{National Astronomical Observatories / Yunnan Astronomical Observatory,
   CAS, Kunming 650011, China
                  email: \url{lyu@ynao.ac.cn}\\
             }

\begin{abstract}
For the future development of Chinese Giant Solar Telescope (CGST)
in Western China, a new sky brightness monitor (SBM) has been
produced for the site survey for CGST. To critically examine the
performance and sensitivity of SBM, we used it in the observation of
the annular solar eclipse in Dali City, Yunnan, on 15 January 2010.
The observation met good weather condition with almost clear sky
during the eclipse. The SBM measurement translates into the solar
illuminance changes at a level of $2.4\times10^{-4}I$ ~s$^{-1}$
during the eclipse. The time of the minimal sky brightness in the
field of view (FOV) is found consistent with the time of maximum
eclipse. Two local sky regions in the FOV are chosen to make time
series of calibrated skylight profiles. The evolution of the sky
brightness thus calibrated also shows good consistency with the
eclipse, particularly between the second and the third contacts. The
minimal sky brightness in each local sky region took place within
half a minute from the corresponding predicted contact time. Such
small time delays were mainly caused by occasional cirri. The
minimal sky brightness measured during the eclipse is a few
millionths of $I_{\odot}$ with standard deviation of 0.11 millionths
of $I_{\odot}$. The observation supports that the single-scattering
process (optically thin conditions) is the main contributor to the
atmospheric scattering. We have demonstrated that many important
aerosol optical parameters can be deduced from our data. We conclude
that the new SBM is a sensitive sky photometer that can be used for
our CGST and coronagraph site surveys.
\end{abstract}
\end{opening}

\section{Introduction}
\label{S-Introduction} The sky brightness is a critical parameter
for judging a potential site for direct coronal observation. At an
excellent coronagraph site, its normalized sky brightness at noon
time is usually around ten millionths of the solar disk center
intensity so that it is possible for coronal emission to be reliably
detected \cite{Lin04,Penn04}. A couple of years ago scientists in
Chinese solar physics society had reached a common understanding
that they would find an excellent solar observation site in Western
China before developing their next-generation large-aperture solar
telescope, the Chinese Giant Solar Telescope (CGST;
\opencite{Fang11}). Based on the experience of the Advanced
Technology Solar Telescope (ATST; \opencite{Keil00}) site survey,
they decided to adopt a sky brightness monitor (SBM) as one of the
necessary tools to precisely measure the Sun's nearby sky brightness
and to provide with information on the precipitable water vapor
content in the atmosphere. It should be noted that before the SBM
was designed by the ATST site survey team, the traditional visual
Evans sky photometer had been used for a long time since 1940's
\cite{Evans48,Lin04}. The SBM is actually a coronagraph instrument
that is able to supply quasi-instantaneous, multi-wavelength,
two-dimensional data with an automatic tracking system and a CCD
system, much more objective, flexible, and powerful than the Evans
sky photometer.

In January 2010 the first CGST SBM was finished in Yunnan and
Nanjing optical labs. It was designed independently but has the same
function as ATST SBMs \cite{Lin04}, and it was equipped with a
16-bit CCD camera (SBIG402ME, 765$\times$510 pixels). Since then,
this new SBM has been used for sky brightness measurements at many
different high-altitude sites of at least 3000m above sea level
(\citeauthor{Liu11a}, \citeyear{Liu11a,Liu11b}). Important
calibration parameters have also been obtained for it. The
instrumental scattered light for different wavelengths are obtained
to be 0.77, 0.81, 2.12, and 3.26 millionths for the blue, green, red
and water vapor bandpasses, respectively. A comparison between the
CGST SBM and the one used in the ATST site survey in
\inlinecite{Lin04} can be found in Table~\ref{T-table1}, which shows
that their instrumental scattered light levels are comparable but
that of the CGST SBM seems slightly better.

\begin{table}
\caption{Comparison of instrumental scattered light level
(normalized in millionth of Sun's disk center intensity). }
\label{T-table1}
\begin{tabular}{ccccc}     
\hline                   
Instrument & Blue & Green & Red & Water vapor \\
\hline
ATST SBM\tabnote{Lin and Penn, 2004} & 1.5 & 1.8 & 3.8 & 3.5  \\
CGST SBM & 0.77 & 0.81 & 2.12 & 3.26   \\

  \hline
\end{tabular}
\end{table}

During the 15 January 2010 annular eclipse, we used our SBM to
measure fast time variations in sky brightness in order to learn
about not only the eclipse itself but also the sensitivity of SBM
which had not been studied or reported before. We will introduce the
observation and data analysis in Sections~\ref{S-observation}
and~\ref{S-analysis} and present the conclusions in
Section~\ref{S-Conclusion}.

\section{Observation} 
\label{S-observation}

\begin{figure}    
\centerline{\includegraphics[width=0.7\textwidth,clip=]{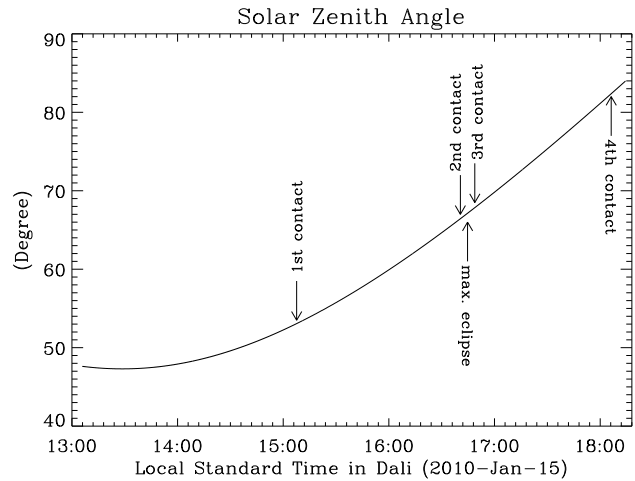}}
\caption{Evolution of the solar zenith angle during the observation. Five vertical arrows show the moments of contact and the maximum eclipse.}
\label{Zangle}
\end{figure}

The observation site ($26^{\circ} 07' 27''$N, $099^{\circ} 57'
15''$E) of the annular eclipse was located near the Cibi lake in
Dali City, Yunnan, at an altitude of 2060 m above sea level. The day
was clear, sunny, and the wind was weak. The eclipse observation was
made in the green wavelength band and with a highest time cadence up
to 1 s. The dark-field and flat-field data had been taken for the
first step correction. A sample image taken with the SBM is shown in
Figure~\ref{SampleEclipse}. The west direction is to the right, and
north to the top. Two arc-shaped sky regions highlighted around the
Sun were the region where local sky brightness was measured. The
inner and outer radii of the two arc-regions are [4.8, 6.1]
$R_{\odot}$ from the solar center, respectively. There are totally
24378 CCD pixels in region I, and 24369 pixels in region II. It
should be noted that the eclipse ended just when the sunset occurred
on the top of the mountains to the west of the observing site.
Therefore, some shadows of a steel tower and the mountains entered
the field of view (FOV) during the late period of the eclipse. In
this case, the pixels in the shadow of the tower will be deleted
from our calculation. From the first contact to the third contact,
the solar zenith angle was between $50^{\circ}$ and $70^{\circ}$
(Figure~\ref{Zangle}), and the eclipse was observed without any
obscuring building or tree in this period.

\begin{table}
\caption{Time table of the observations. } \label{T-table2}
\begin{tabular}{ccccc}     
\hline                   
No. & Time & Description \\
  \hline
1 & 15:07:37 & First contact; start of the eclipse   \\
\\
2 & 16:40:36 & Second contact   \\
3 & 16:40:46 & Min. brightness in sky region I \\
\\

4 & 16:44:42 & Min. brightness in full sky \\
5 & 16:44:42 & Max. eclipse    \\
\\

6 & 16:48:24 & Min. brightness in sky region II \\
7 & 16:48:48 & Third contact   \\
\\
8 & 18:06:16 & Fourth contact; end of the eclipse   \\
\hline
\end{tabular}
\end{table}

The eclipse started at 15:07:37 (local standard time; hereinafter
the same) when the Moon was observed transiting into the Sun's disk
from its southwest limb. The eclipse ended at 18:06:16 and lasted
for almost 3 h. Table~\ref{T-table2} lists the key times for this
annular eclipse. According to NASA's report
(http://eclipse.gsfc.nasa.gov/ SEcat5/ SE2001-2100.html), the
magnitude of this annular eclipse was 0.919, and it was the longest
annular solar eclipse in this century.

\section{Data Analysis}
\label{S-analysis}
\subsection{Calibration of Sky Brightness} 
\label{S-calibration}

Usually, the direct sky brightness measurements should be calibrated
by normalizing each sky data with respect to the solar disk center
intensity. However, it is not so straightforward in the case of an
eclipse in which the solar disk center is sometimes partially and
sometime totally covered by the Moon.

In our observation, in order to make a high-precision sky intensity
measurement, the exposure time was preset so that the sky brightness
value could fill approximately two-thirds of the full-well capacity
(corresponding to 65535) of the 16-bit CCD. In this way the Sun's
disk pixels, including both limb and center areas, had to become
overexposed for most of the observing time. Only after 16:26 when
the solar zenith angle reached $60^{\circ}$ the readout of the limb
pixels (at 0.9 radii from the disk center) dropped to below 60000.

\begin{figure}    
\centerline{\includegraphics[width=1.\textwidth,clip=]{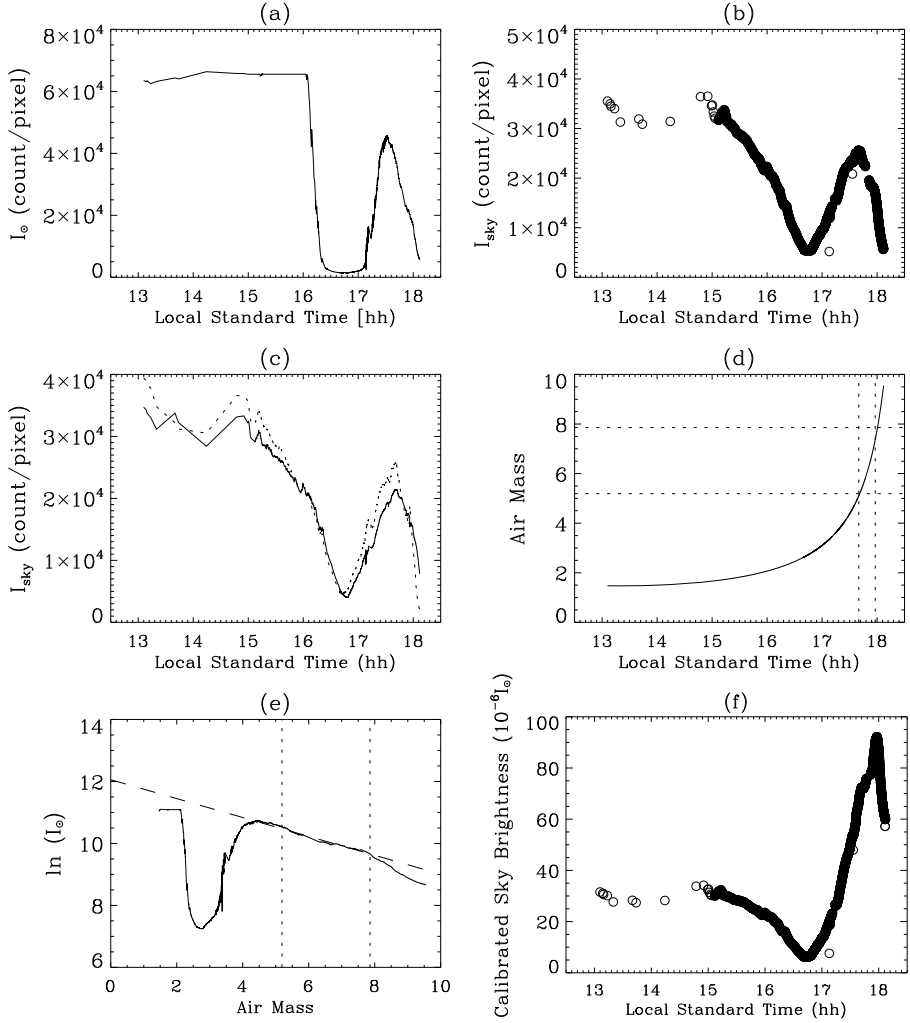}}
\caption{(a) Time profile of un-calibrated $I_{\odot}$. (b) Time
profile of un-calibrated $I_{\rm sky}$. (c) Two time profiles of
calibrated $I_{\rm sky}$ for regions I (dashed curve) and II (solid
curve), respectively. (d) Time profile of air mass. (e) Profile of
ln$(I_{\odot})$ as a function of air mass. The data between the two
vertical lines are used to extrapolate ln$(I_{\odot})$ as is shown
with the inclined dashed line. (f) Time profile of the  calibrated
sky brightness.} \label{AirMass}
\end{figure}

Figure~\ref{AirMass}(a) shows raw measurements of the solar disk
intensity for most of the time of this annular eclipse, with the
solar disk either saturated due to CCD overexposure (before
$\sim$16:05) or occulted by the Moon ($\approx$16:05-17:30).
Figure~\ref{AirMass}(b) shows the variations in the sky illuminance
(directly from the CCD readout) averaged in the sky region [4.8,
6.1] $R_{\odot}$. There is an obvious flux drop during the eclipse
time. Moreover, the two flux profiles in Figure~\ref{AirMass}(c)
show the variations in the sky illuminance averaged over the local
regions I and II, respectively. Their profiles are similar to that
of the whole sky area in Figure~\ref{AirMass}(b). Unfortunately,
such sky brightness data shown in Figure~\ref{AirMass}(b) and
\ref{AirMass}(c) are not useful since no corrections for the
atmospheric extinction have been made. That is, the time evolution
of the sky brightness must be calibrated for further quantitative
study.

When checking the data after 17:40 (Figure~\ref{AirMass}(a)), we
found that the central part of the solar disk was not only exposed
well but also already outside of the lunar shadow. Thus, the
normalized sky brightness data between 17:40 and 18:07 (which was
before the end of the observation) can supply us the opportunity to
deduce the atmosphere thickness which is critical for the next
calculations of solar disk center intensity for the other times.

Our calculation of air mass is based on a uniform curved atmospheric
model as a function of zenith angle~\cite{Lin04}.
Figure~\ref{AirMass}(d) presents the time evolution of our air mass
model. From the theory of atmospheric extinction, the logarithm of
the solar disk center intensity should be proportional to the air
mass for constant extinction in cloud-free conditions~\cite{Penn04}.
We plot the profile of $\ln(I_{\odot})$ as a function of air mass in
Figure~\ref{AirMass}(e), finding a nearly linear relation in the
period between 17:40:00 and 17:58:00. This period is also shown
between the vertical dashed lines in Figure~\ref{AirMass}(d). The
corresponding range in the air mass [5.19, 7.86] is shown between
the two vertical dashed lines in Figure~\ref{AirMass}(e). With the
linear relationship between $\ln(I_{\odot})$ and air mass for
cloud-free conditions, the theoretical solar disk center data for
the other times before 17:40:00 can be easily obtained by the
extrapolation from the data between 17:40:00 and 17:58:00. The
results of the fitting and extrapolation for $\ln(I_{\odot})$ as a
function of air mass are shown by the dashed line in
Figure~\ref{AirMass}(e). The extinction coefficient derived from the
slope of the fitted line is 0.304.

Finally, by normalizing to the extrapolated solar disk center intensity and after removing the instrumental scattered light (Table~\ref{T-table1}), the time series of calibrated sky
brightness have been obtained (Figure~\ref{AirMass}(f)). After the third contact the sky brightness kept increasing obviously, but at about 17:55 the CCD data in region I started to decrease significantly due to the shadow of the west mountains in this sky region. Under this condition we use the pixels in another clear sky region. At about 18:07:03 we stopped the observation when it was only a few seconds before the total sunset. Therefore, at the fourth contact at 18:06:16 when the Sun was not obscured any more by the Moon, the sky was already dark.

It should be noted that the sky brightness measurements started at
13:05, about 2 h before the first contact, but with a random time
cadence and relatively scarce data samples. From the limited samples
before the eclipse, we can still see that the average sky brightness
kept at a stable level of about 27-34 millionths of $I_{\odot}$,
indicating good atmospheric condition on the eclipse day which was
suitable for testing a photometer.

\subsection{Study of Sky Brightness} 
\label{S-results}

\begin{figure}    
\centerline{\includegraphics[width=0.9\textwidth,clip=]{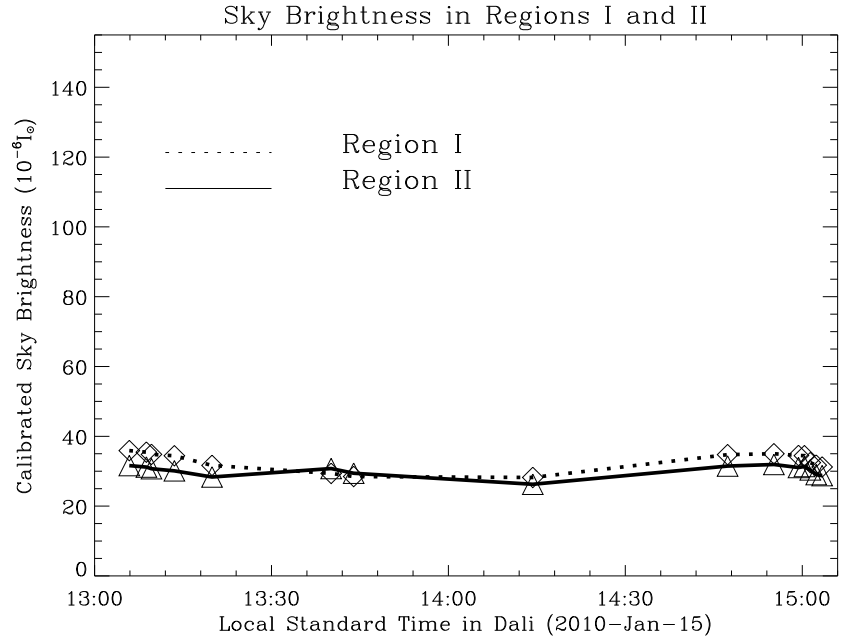}}
\caption{Time profiles of calibrated sky brightness for the two regions I (thick dotted line) and II (thick solid line) before the start of the annular eclipse.}
\label{FluxBeforeEclipse}
\end{figure}

Figure~\ref{FluxBeforeEclipse} shows the calibrated profiles of the
sky brightness before the eclipse for the two local regions I and
II, respectively. Small-amplitude and slow variations can be
noticed, which should reflect normal air turbulence due to, {\it
e.g.}, occasional cirri. It can be found that before the eclipse,
the difference in sky brightness between regions I and II was always
within 5 millionths of $I_{\odot}$, but region I showed higher
values for most of the time. The main reason, we suppose, should be
an uneven distribution of the scattered light in SBM. For the
purpose of greatly reducing the instrumental scattered light, a set
of O-rings have been used and stacked along the cylindrical
container of neutral density filters (ND4; $10^{-4}$ transmission)
in front of the SBM tube. We find from the original data series
({\it e.g.}, Figure\ref{SampleEclipse}) that some stronger scattered
light existed at the immediate west edges of the O-rings, just close
to region I. Usually, such an uneven scattered light distribution is
unnoticeable in the flat-field data. In any case, the intensity of
scattered light is proportional to the incident light, and the
difference in the scattered light levels in the two regions is
small, so that it should not affect our examination of the
sensitivity of the instrumental response to sky illuminance changes
during the eclipse.

\begin{figure}    
\centerline{\includegraphics[width=0.9\textwidth,clip=]{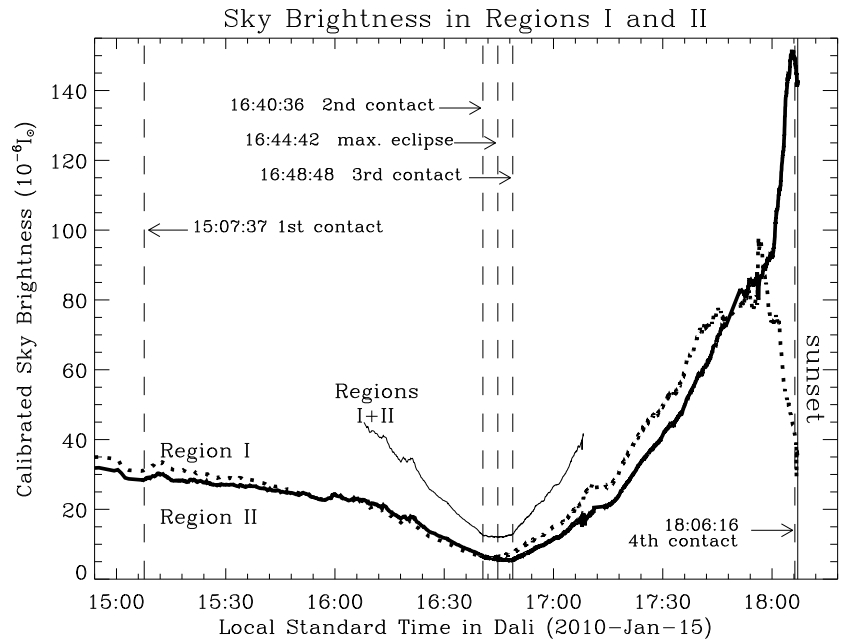}}
\caption{Time profiles of calibrated sky brightness for regions I (dotted line), II (thick solid line) and I+II (thin solid line) during the annular eclipse.}
\label{FluxDuringEclipse}
\end{figure}

Figure~\ref{FluxDuringEclipse} presents the calibrated sky
brightness during the eclipse for regions I and II, and the
combination of I plus II, respectively. The sky brightness in
regions I and II became nearly the same before the second contact of
the eclipse. For the sky region of I plus II, shown with a thin
solid line for one hour including the maximum eclipse time, the sky
brightness showed a profile almost symmetric in time during this
period. The minimum values of the three profiles in
Figure~\ref{FluxDuringEclipse} are all found to occur in the period
from the second to the third contact. In Table~\ref{T-table2}, we
list the key times for the annular eclipse and the brightness
measurements in different sky regions. From this comparison, good
time correspondence is found between the contacts and their
accompanying sky brightness changes.  Within the measurement
accuracy, we find no time difference among the maximum eclipse and
the minimum brightness in the FOV. This good temporal coincidence
indicates that our SBM is a sensitive photometer that can track
rapid changes in the sky brightness during the eclipse.

Moreover, in region I the sky brightness dropped to the minimum 10 s
after the time of the second contact, while in region II it occurred
24 s before the time of the third contact. These small time
differences between the observed minima in sky brightness and the
theoretical expectations may indicate some weak atmospheric
turbulence due to dust, cloud, or other components that are not
easily noticed by eye but can be detected by the SBM instrument.
There was a 7.5 min difference between the sky brightness minima in
regions I and II. This time difference, almost equal to the time
between the second and the third contact, suggests that the
atmospheric scattering process in each sky region should be
independent from each other in some degree, which we will try to
study in the next section.

\subsection{Comparison with Simulation and Theory} 
\label{S-comparisons}

For comparison, we have carried out a simulation of the annular
eclipse and calculated its normalized illuminance above the
terrestrial atmosphere (Figure~\ref{Simulation}(a-b)), based on the
eclipse magnitude and the limb darkening
parameters~\cite{Livingston00}. An animation movie from the
simulation is available in the online journal. The normalized
illuminance profile is shown in Figure~\ref{Simulation}(b), with the
minimum value of 0.108. Figure~\ref{Simulation}(c) shows a
photograph of the annular eclipse taken at the maximum obscuration.
In Figures~\ref{AirMass}(f),~\ref{FluxDuringEclipse},
and~\ref{Simulation}, a flat-bottom feature of about 8 min width can
be noticed in the valley part of the profiles around the eclipse
maximum. A close-up plot including this feature is shown in
Figure~\ref{Flux_I_II} over a shorter period including the second
and the third contacts.

The profile of the normalized solar illuminance obtained in the
simulation, multiplied by a factor of 70, is also shown as the
dashed line in Figure~\ref{Flux_I_II}. The thin solid line
represents the profile of the combination area of I plus II after
shifting down by 4.46, {\it i.e.}, $B_{\rm I}+B_{\rm II}-4.46$. It
can be clearly seen that the two profiles fit well with each other
during the period from the second contact to the third contact. The
standard deviation between the two profiles over this period is only
0.11 millionths of $I_{\odot}$, suggesting high sensitivity of our
SBM instrument to weak signals.

The two arrows in Figure~\ref{Flux_I_II} denote the minima in the
profiles for regions I and II, respectively. The minimum of region I
was obviously earlier than that of region II. In the eclipse
observation, the Moon started to transit from the southwest side of
the solar disk. That is, the place for the first two contacts to
occur was closer to region I than to region II, indicating that the
single-scattering process (optically thin conditions) should be the
main contributor to the atmospheric scattering.

In the following, the classical theory on sky radiation under the
condition of single scattering is used to deduce the sky brightness
in a homogeneous atmosphere~\cite{Gonzalez98, Lin04,Huang08}.
According to those theories the normalized sky brightness at a
wavelength $\lambda$ along the line of sight (LOS) is:
\begin{eqnarray}
B_{\lambda}(\hat{\bf n}) &=& \frac{I_{\lambda}(\hat{\bf n})} {I_{\lambda}(\hat{\bf n}_{\odot})} \nonumber\\
&=& \Phi_{\lambda}(\hat{\bf n}) \cdot\kappa_{\lambda}\cdot M(\hat{\bf n})+B'_{\lambda} \nonumber \\
&\simeq& \Phi_{\lambda}(\hat{\bf n}) \cdot\kappa_{\lambda}\cdot
M(Z_{\odot}) + B'_{\lambda} \,,
\label{Eq-scatter-func}
\end{eqnarray}
where $\hat{\bf n}$ is the unit vector along LOS, \ $\hat{\bf
n}_{\odot}$ the unit vector from the observer to the solar disk
center, and $\kappa_{\lambda}$ the extinction coefficient.
$M(Z_{\odot})$ is the air mass along the direction of the Sun, and
$Z_{\odot}$ is the zenith angle, {\it i.e.}, the zenith distance of
the Sun. In our observation, since the sky regions were close to the
Sun, it can be considered that $M(\hat{\bf n})\simeq M(Z_{\odot})$.
$B'_{\lambda}$ is an instrumental constant which depends on the
wavelength. $\Phi$ is the angular (scattering) phase function,
describing the fall of the sky brightness distribution when moving
away from the solar limb, which can be expressed as follows:

\begin{eqnarray}
\Phi_{\lambda}(\hat{\bf n}) &=& \frac{1}{4 \pi} \int_{\Delta
\omega_{\hat{\bf n}_{\odot }} } P(\hat{\bf n},\hat{\bf n}')
\frac{I^{0}_{\lambda}(\hat{\bf n}')} {I^{0}_{\lambda}(\hat{\bf
n}_{\odot})} \rmd \omega_{\hat{\bf n}'} \,, \label{Eq-phi}
\end{eqnarray}
where $\Delta \omega_{\hat{\bf n}_{\odot }}$ is the solid angle
subtended by the solar disk as viewed from the observer in the
terrestrial atmosphere. $I^{0}_{\lambda}(\hat{\bf n}$) is the solar
illumination function outside the terrestrial atmosphere, and it
actually provides the center-to-limb variation. In the visible
wavelength range it is the well-known limb darkening function. $P$,
the so-called adequate phase
function~\cite{Chandrasekhar50,Pillet90} describing the scattering
process, can be practically given by the Henyey-Greenstein phase
function~\cite{Henyey41}:
\begin{eqnarray}
P &=& \frac{1-g^{2}} {\left[ 1+g^{2}-2g \cos\theta\right]^{3/2}} \,,
\label{Eq-P}
\end{eqnarray}
in which,
\begin{eqnarray}
\cos\theta &=& \hat{\bf n} \cdot \hat{\bf n}' \,,
\end{eqnarray}
and the free parameter ($g$) gives the angular dependence of the
scattered light with $g=1,0,-1$ for forward, isotropic, and backward
scattering, respectively. From our simulation based on
Equation~(\ref{Eq-P}), $P$ is found sensitive to the scattering
angle factor $\cos\theta$ for a given value of $g$.

On the other hand, for a given direction $\hat{\bf n}$,  $g$ (thus,
the angular phase function $\Phi$) and $M$ can be thought constant
during a short time ({\it e.g.}, from the second contact to the
third). The evolution of the sky brightness in this direction shall
be dependent on the solar radiation function in the same direction.
During an eclipse the Moon just plays the role to `cut' the direct
solar radiation from the space, resulting in a zero solar radiation
function which will significantly reduce the sky brightness in its
shadow, accordingly. Therefore, it is not difficult to understand
there is a time difference (7.5 min) for the minima in regions I and
II. Moreover, the non-zero sky brightness for the minimum in each
sky region should be mainly contributed by the atmospheric
scattering from the other part of the solar disk that was not
occulted by the Moon. We expect a near zero sky brightness minimum
for a total solar eclipse.

\section{Discussion and Conclusions}
\label{S-Conclusion} For the purpose of CGST and coronagraph site
survey in Western China, we had recently developed a new SBM
following the ATST team's experience. During the Dali annular
eclipse on 15 January 2010, we used it to measure the sky brightness
to test its performance and sensitivity. The sky was clear and sunny
on that day, and the eclipse ended just before the time of the local
sunset.

The calibration of the sky brightness data is special for the case
of the eclipse. In order to obtain the theoretical solar disk center
intensity, we used the extrapolation method \cite{Penn04} and the
uniform curved atmospheric model \cite{Lin04} based on the data
available between 17:40 and 18:07. Because of the nice atmospheric
condition during the observation, the air mass and the theoretical
solar intensity data were successfully obtained from the short-term
observation.

\begin{figure}    
\centerline{\includegraphics[width=0.7\textwidth,clip=]{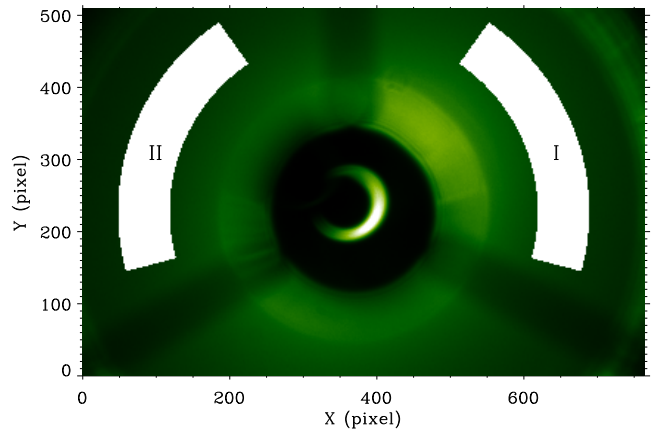}}
\caption{A sample eclipse image taken by SBM in the green band. Two arc-shaped sky regions, marked with `I' and `II', are the regions where we analyzed the sky brightness in detail. The central region shows the eclipsed solar disk filtered by the neutral density occulter. The shadows are due to three occulter support arms. Outside of this occulter region are bright diffraction rings from the occulter edges. Diffraction from the front of the telescope tube can be seen at the rightmost edges of the field of view.}
\label{SampleEclipse}
\end{figure}

We have calculated the sky brightness for the FOV and two local sky
regions around the solar disk (Figure\ref{SampleEclipse}). These sky
brightness profiles show a common, flat-valley feature between the
second and the third contacts. The noise signal due to local
atmospheric turbulence was clearly seen through the comparison
between the observation and the simulation (see the upper two lines
in Figure~\ref{Flux_I_II}). It is interesting to find that the times
of their minimum showed good agreement with the eclipse stages. The
brightness minimum of the whole FOV occurred just at the same time
of the maximum eclipse (Table~\ref{T-table2}). The minimum in the
sky brightness took place 10 s later (24 s earlier) than the second
(third) contact for region I (II) which is at the west (east) side
of the solar disk, respectively. These two time differences,
although small, suggest us that it is better to calculate the sky
brightness using the sky area as large as possible in the FOV for
the site survey work. As a matter of fact we found no time delay for
the whole FOV data.

On the other hand, we have noticed a 7.5 min time difference between
the sky brightness minima in regions I and II. The time difference
actually corresponds to the period from the second to the third
contact of the eclipse. This phenomenon supports that single
scattering is the main scattering process in the terrestrial
atmosphere, and multiple scattering can be ignored for optically
thin conditions.

Moreover, we have noticed that the lowest sky brightness values
obtained by the SBM were only a few millionths of $I_{\odot}$ for
this eclipse between the second and the third contacts
(Figure\ref{Flux_I_II}). We can use these measured low sky
brightness, together with other parameters from the calculations and
simulations as mentioned above, to further investigate the
scattering process in the atmosphere. One of the advantages of the
15 January 2010 solar annular eclipse, is that it offered us the
longest time for a high-precision measurement of the sky brightness
from the second contact to the third one. Here, based on some
reasonable assumptions, we will deduce and estimate the free
parameter $g$ in the Henyey-Greenstein phase function in
Equation~(\ref{Eq-P}). The angular dependence of the scattered light
is determined by $g$; therefore it is an important parameter for
diagnosing optical properties of aerosol.

Since the observed regions I and II were close to the Sun, we can
simplify the Henyey-Greenstein phase function to be:
\begin{eqnarray}
P &=& \frac{1+g} {\left( 1-g\right)^{2}}   \,, \label{Eq-P2}
\end{eqnarray}
with $\cos\theta\simeq1$. On the other hand, based on Equations~(\ref{Eq-scatter-func}) and (\ref{Eq-phi}), $P$ can be expressed by:
\begin{eqnarray}
P &=& \frac{4\pi B L^{2}} { (1-m ^{2}) \kappa_{\lambda} M I_{\lambda}^{1}}
\,, \label{Eq-P3}
\end{eqnarray}
in which all the physical and geometrical parameters are
nondimensional numbers. That is, $B$ is the calibrated sky
brightness, $L$ is the Sun-Earth distance in unit of solar radius,
$m$ is the magnitude of the annular eclipse, $\kappa_{\lambda}$ is
the extinction coefficient, $M$ is the air mass, and
$I_{\lambda}^{1}$ is the normalized solar illuminance. Because the
parameters on the right-hand side of Equation~(\ref{Eq-P3}) have
been known from the last sections for the time of the maximum
eclipse (16:44:42), $g$ can be solved out from
Equations~(\ref{Eq-P2}) and~(\ref{Eq-P3}). The result and the
parameters involved in the calculation are listed in
Table~\ref{T-table3}. The value of $g$, 0.912, indicates a strong
forward-scattering feature in the atmosphere for Dali city, Yunnan.

\begin{table}
\caption{Nondimensional parameters for atmospheric optical
properties derived at the maximum eclipse. See the text for
definitions} \label{T-table3}
\begin{tabular}{cccccccc}     
\hline                   
Time     & $B$                 & $L$ & $m$ & $\kappa_{\lambda}$ & $M$& $I_{\lambda}^{1}$ & $g$\\
\hline
16:44:42 & $6.22\times10^{-6}$ &215.1&0.919 & 0.304& 2.53 &0.108 & 0.912  \\
\hline
\end{tabular}
\end{table}

\begin{figure}    
\centerline{\includegraphics[width=0.9\textwidth,clip=]{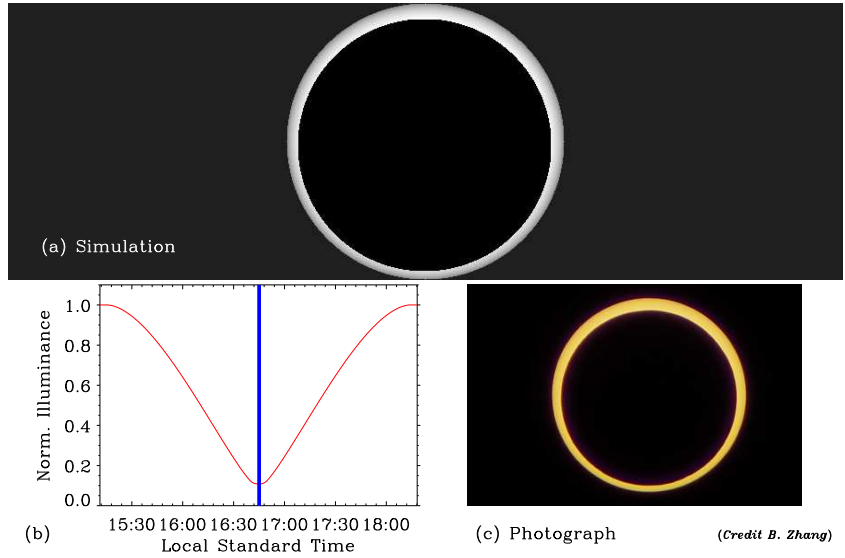}}
\caption{(a) Simulation of the annular eclipse. An animation file is available in the online journal. (b) Normalized solar illuminance during the eclipse based on the simulation. (c) A sample eclipse photograph.}
\label{Simulation}
\end{figure}

\begin{figure}    
\centerline{\includegraphics[width=0.9\textwidth,clip=]{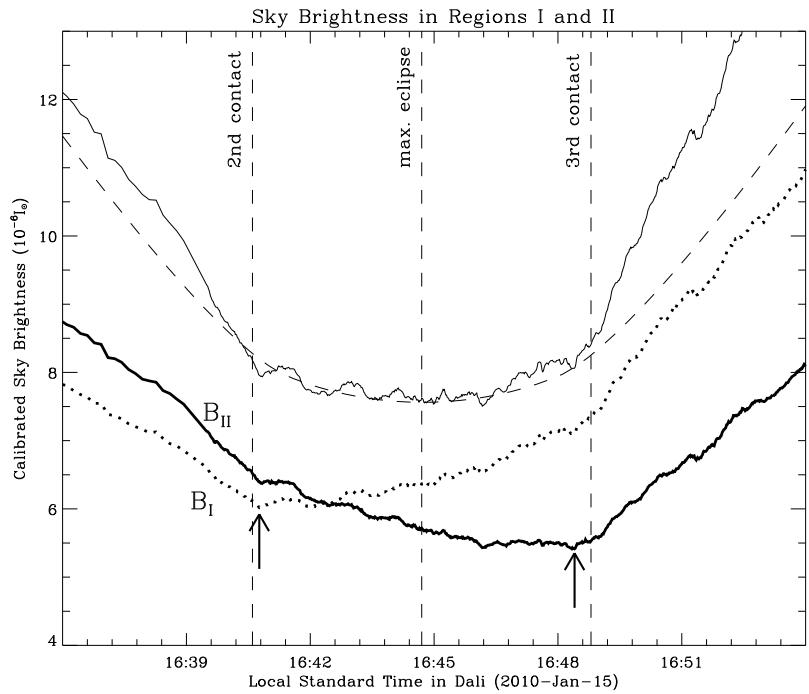}}
\caption{Time profiles of calibrated sky brightness for regions I
(thick dotted line), II (thick solid line) and I+II (thin solid
line, shifted down along the ordinate; see text) during the annular
eclipse. The dashed line represents the normalized solar illuminance
simulated in Figure~\ref{Simulation} (multiplied by a factor; see
text). The two arrows point to the minimum sky brightness in regions
I and II.} \label{Flux_I_II}
\end{figure}

If we adopt a simple model of plane-parallel atmosphere, our SBM
measurement can be translated into the changes in solar illuminance
as small as $2.4\times10^{-4}I~{\rm s}^{-1}$ (from
Figure~\ref{Simulation}) during the eclipse, where {\it I} is the
Sun's illuminance without any obscuration by the Moon. For
comparison, a large impulsive X-ray flare can contribute to the
solar illuminance increase in the whole visible spectral region like
$1.1\times10^{-4}I$~{\rm s}$^{-1}$ \cite{Svestka76}. Thus we are
satisfied with the SBM sensitivity.

In 2011 US National Solar Observatory had shipped their SBM system
to us for calibrating the CGST SBM. The calibration data have been
obtained and further calibration results will be shown in another
paper.

There will be another annular eclipse to occur soon on 21 May 2012
in southeast of China. The difference is that this eclipse will
happen near sunrise. We need to preset proper exposure times for the
SBM to measure the brightness of the sky and the photosphere
simultaneously in all four wavelengths, in order to reveal more
information about the scattering process in the atmosphere.

\begin{acks}
The authors thank the referee and the editor for many useful
suggestions which improved the manuscript. This work has obtained
encouragements and assistance from many colleagues. We thank
especially Hai-sheng Ji, Hui Li, Bai-rong Zhang, Ming-chan Wu, Ke
Lou, Zhong Liu, Jun Lin, Lei Yang, Hou-kun Ni and Haiying Zhang for
discussion and support. This work is supported by the Natural
Science Foundation of China (Grant Nos. 10933003, 11078004, and
11073050) and the National Key Research Science Foundation (MOST
2011CB811400).
\end{acks}

\clearpage

\end{article}

\end{document}